\documentclass[11pt]{article}
\usepackage{amsfonts,amsmath,amssymb,epsfig,float,here,latexsym,setspace}
\usepackage[all]{xy}
\usepackage[square, comma]{natbib}
\usepackage[english]{babel}
\usepackage{hyperref}
\usepackage[page]{appendix}

\def\qed{\unskip\nobreak\hfill$\Box$\par\addvspace{\medskipamount}}

\newcommand{\be}{\begin{equation}}
\newcommand{\ee}{\end{equation}}
\newcommand{\bea}{\begin{eqnarray}}
\newcommand{\eea}{\end{eqnarray}}
\newcommand{\beas}{\begin{eqnarray*}}
\newcommand{\eeas}{\end{eqnarray*}}

\newtheorem{theorem}{Theorem}[section]
\newtheorem{definition}[theorem]{Definition}

\newtheorem{remark}[theorem]{Remark}
\newtheorem{example}[theorem]{Example}
\newtheorem{examples}[theorem]{Examples}
\newtheorem{foo}[theorem]{Remarks}

\begin{document}
\title{\vskip -0.4cm
Probability Premium and Attitude Towards Probability
}
\author{Louis R. Eeckhoudt\\
{\footnotesize I\'ESEG School of Management}\\
{\footnotesize Catholic University of Lille}\\
{\footnotesize and CORE}\\
{\footnotesize {\tt Louis.Eeckhoudt@fucam.ac.be}}\\\and Roger J. A. Laeven\thanks{Corresponding author.
Mailing address: University of Amsterdam, Amsterdam School of Economics, PO Box 15867, 1001 NJ Amsterdam, The Netherlands.
Phone: +31 20 525 4219/4252.}\\
{\footnotesize Amsterdam School of Economics}\\
{\footnotesize University of Amsterdam, EURANDOM}\\
{\footnotesize and CentER}\\
{\footnotesize {\tt R.J.A.Laeven@uva.nl}}\\
[0.0cm]}
\date{This Version:
\today} \maketitle
\begin{abstract}
Employing a 
generalized definition 
of Pratt \cite{P64} and Arrow's \cite{A65,A71}
probability premium,
we introduce a new concept of attitude towards probability.
We illustrate in a problem of risk sharing that whether attitude towards probability is a first-order or second-order phenomenon
has 
important 
economic applications. 
By developing a local approximation to the probability premium,
we show that the canonical rank-dependent utility model usually exhibits attitude towards probability of first order,
whereas under the dual theory with smooth probability weighting functions attitude towards probability is a second-order trait. 
\noindent
\\[4mm]\noindent\textbf{Keywords:}
Probability Premium; Expected Utility; Dual Theory; Rank-Dependent Utility;
Local Risk Aversion; Risk Sharing.
\\[4mm]\noindent\textbf{AMS 2010 Classification:} Primary: 91B06, 91B16, 91B30; Secondary: 60E15, 62P05.
\\[4mm]\noindent\textbf{JEL Classification:} D81, G10, G20.
\end{abstract}

\makeatletter
\makeatother
\maketitle

\newpage

\onehalfspacing

\setcounter{equation}{0}

\section{Introduction}\label{sec:intro}

Under expected utility (EU), there exist various intimately related 
measures
to evaluate the degree of risk aversion of a decision-maker (DM).
These measures include the risk premium, the probability premium and the local index of absolute risk aversion, originally developed independently
by Pratt \cite{P64} and Arrow \cite{A65,A71} a little more than 50 years ago.\footnote{See also the early contribution by 
de Finetti \cite{dF52} (written in Italian), which already contains the local index of absolute risk aversion.} 
Since then they have played highly versatile 
and important roles in the theoretical and experimental analysis of risky choices under EU
and have seen widespread applications across many fields.

Among these measures and despite its appealing nature, the probability premium 
has for a long time been by far the least popular. 
Only in recent years it starts receiving increased attention 
(Jindapon \cite{J10}, Liu and Meyer \cite{LM13}, Eeckhoudt and Laeven \cite{EL15}, and Liu and Neilson \cite{LN19}).
The probability premium, 
which contrary to the risk premium has been developed only for a binary symmetric zero-mean risk,
may be given a simple 
interpretation:
it answers the question of by how much the probability of the adverse payoff in a binary zero-mean risk needs to be reduced---and
the probability of the prosperous payoff needs to be increased accordingly---,
to make the DM indifferent between bearing and not bearing the risk.

In this paper, we start by generalizing Pratt's \cite{P64} definition of the probability premium such that it becomes
applicable also to mean-preserving risk changes with probability mass less than unity,
while maintaining its simple interpretation.\footnote{Liu and Neilson \cite{LN15,LN19}
generalize Pratt's definition of the probability premium in another direction,
to allow for random starting wealth and to risk aversion of higher degrees.}
Next, we employ this general 
definition of the probability premium 
to introduce a new concept of attitude towards probability.
We say that a DM displays attitude towards probability of order 1 and order 2 if,
for risk changes with small probability mass,
the associated probability premium is linear in the probability mass and proportional to the square of the probability mass, respectively.
These orders of attitude towards probability can be seen to represent the ``probability counterparts''
of the well-known orders of attitude towards risk introduced by Segal and Spivak \cite{SS90}.

We show that whether attitude towards probability is a first- or second-order phenomenon has 
important economic applications.
We illustrate in particular the 
implications of first- and second-order probability aversion
in a problem of risk sharing.
We reveal that a first-order probability averse DM facing a risk with a small loss probability may opt for 
risk sharing among individuals
even when the other individuals' loss probability is 
larger than the individual's own loss probability.
By contrast, a second-order probability averse individual facing the same risk engages only in actuarially fair 
risk sharing.
Our results on risk sharing can be viewed as dual to the well-known Mossin's \cite{M68} Theorem
on the sub-optimality of full insurance in the presence of a positive safety loading.
In Mossin's Theorem there is a trade-off between the degree of risk aversion
and a positive safety loading in the ``payoff plane''.
In our result, there is a trade-off between the degree of probability aversion
and actuarially unfair loss probabilities
in the ``probability plane''.

Measures of risk aversion have also received much attention outside EU.
In particular, in the dual theory (DT; Yaari \cite{Y87}) and under rank-dependent utility (RDU; Quiggin \cite{Q82}),
a rich literature
has proposed several measures of risk aversion
and has established the associated results on comparative risk aversion;
see Yaari \cite{Y86,Y87}, Chew, Karni and Safra \cite{CKS87}, Ro\"ell \cite{R87}, Chateauneuf, Cohen and Meilijson \cite{CCM04}, Ryan \cite{R06} and Eeckhoudt and Laeven \cite{EL20}.
This does, however, not apply to the probability premium,
which has almost never been considered in non-EU.\footnote{To our best knowledge, the only exception is the note by Eeckhoudt and Laeven \cite{EL15} providing graphical illustrations.}
As a contribution of independent interest, we provide a local approximation to the probability premium under RDU
and establish the corresponding global results.
As is well-known,
RDU encompasses EU and DT as special cases
and is the main building block of (cumulative) prospect theory (Tversky and Kahneman \cite{TK92}).\footnote{See Schmidt and Zank \cite{SZ08} for global measures of risk aversion
under prospect theory.
}

We exploit our local approximation to the probability premium to show that
RDU and EU DMs usually exhibit first-order attitude towards probability, 
whereas a DT DM with smooth probability weighting function displays second-order probability attitude.
Our results apply to quite general utility and probability weighting functions under RDU.
In particular, we allow for inverse $s$-shaped probability weighting functions
such as those in Prelec \cite{P98} and Wu and Gonzalez \cite{WG96},
which are relevant for descriptive purposes (Abdellaoui \cite{A00}).
That is, we allow for violations of e.g., global risk aversion.

Our paper is organized as follows.
In Section \ref{sec:probpremium} we provide our generalized definition of the probability premium.
In Section \ref{sec:probaversion} we introduce first- and second-order probability aversion.
Section \ref{sec:risksharing} applies first- and second-order probability aversion to explore a risk sharing problem.
In Section \ref{sec:rdu} we develop our local approximation to the probability premium under RDU,
analyze the degree of probability aversion under RDU,
and establish the corresponding global results.
Proofs are relegated to Appendix \ref{sec:proofs}.
Appendix \ref{sec:nonbinary} generalizes the definition of the probability premium
and our local approximation results to cover non-binary risks.

\setcounter{equation}{0}

\section{Probability Premium}\label{sec:probpremium}

\begin{figure}[H]
\begin{center}
\caption{Risk $C$
}
\vskip 0.4cm
\includegraphics[scale=1.40,angle=0]{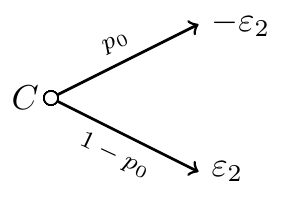}
\label{fig:C}
\end{center}
\end{figure}

\begin{figure}[H]
\begin{center}
\caption{Risk $D$
}
\vskip 0.4cm
\includegraphics[scale=1.40,angle=0]{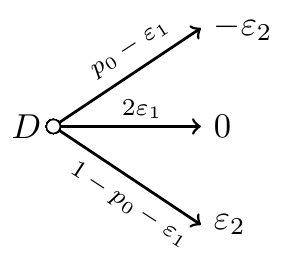}
\label{fig:D}
\end{center}
\end{figure}

We start by considering a risk $C$
yielding outcomes $-\varepsilon_{2}$ and $\varepsilon_{2}$ with probabilities $p_{0}$ and $1-p_{0}$, respectively,
as represented in Figure \ref{fig:C}:\footnote{Henceforth, in all figures,
values alongside the arrows represent probabilities,
while values at the end of the arrows represent outcomes.}$^{,}$\footnote{We assume $0<p_{0}<1$ and $\varepsilon_{2}>0$.}
\vskip -0.5cm
\noindent Risk $C$ is faced by a DM with initial wealth level $w_{0}$.
We transform risk $C$ into a risk $D$ given by Figure \ref{fig:D}:\footnote{We assume $0<\varepsilon_{1}<\min\{p_{0},1-p_{0}\}$.}
\vskip -0.5cm
\noindent We obtain $D$ from $C$ by moving a probability $\varepsilon_{1}$
from both states of $C$ to a new intermediate state with outcome $0$.
Of course, $D$ is a mean-preserving contraction of $C$ (in the sense of Rothschild and Stiglitz \cite{RS70}),
and is thus viewed as an improvement by any (strongly) risk-averse DM.\footnote{Conventionally, risk aversion is often defined as
a preference for a sure wealth level
to a risky wealth position with the same expectation.
This definition of risk aversion is sometimes referred to as ``weak risk aversion''.
A stronger definition of risk aversion consists in requiring a preference for mean-preserving contractions
(Rothschild and Stiglitz \cite{RS70}) of payoff distributions,
also when the resulting payoff distribution is non-degenerate.
Yet another, more elementary, definition of risk aversion
requires that preferences are payoff (or outcome) convex (Yaari \cite{Y69}).
Under EU, all three definitions agree, but this is no longer the case in non-EU;
see e.g., Chew and Mao \cite{CM95}, Cohen \cite{C95}, Nau \cite{N03} and Machina \cite{M13}.
We will henceforth adopt the second definition of risk aversion that draws upon Rothschild and Stiglitz \cite{RS70}.}

\begin{figure}[H]
\begin{center}
\caption{Risk $C(\mu)$
}
\vskip 0.4cm
\includegraphics[scale=1.40,angle=0]{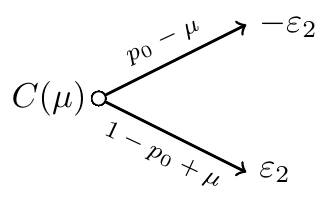}
\label{fig:Cmu}
\end{center}
\end{figure}

We can adapt the probability of the occurrence of the unfavorable state in $C$
from $p_{0}$ to $p_{0}-\mu$,
and adapt the probability of the occurrence of the favorable state in $C$
from $1-p_{0}$ to $1-p_{0}+\mu$ accordingly,
such that the DM becomes indifferent between the resulting risk $C(\mu)$ represented in Figure \ref{fig:Cmu}
and the risk $D$ of Figure \ref{fig:D}.
\vskip -0.5cm
\noindent (Of course, $C(0)\equiv C$.)
Naturally the difference, $\mu$, between the initial probability of the unfavorable state, $p_{0}$,
and the adapted probability of the unfavorable state, $p_{0}-\mu$,
represents the probability premium
corresponding to the risk change from $C$ to $D$.
The probability premium may be positive or negative
depending on whether the DM is risk averse or risk seeking.

From the perspective of income inequality, the transformation from $C$ to $D$ may be interpreted to reduce the proportion of rich and poor people,
and to increase the proportion of middle class people.
The probability premium may then be interpreted as the increase in the proportion of rich people
(and reduction of the proportion of poor people accordingly) in $C$ that makes it equivalent to $D$.

\begin{figure}[H]
\begin{center}
\caption{Risk $C(\mu)$ after omitting the common components of $C(\mu)$ and $D$
}
\vskip 0.4cm
\includegraphics[scale=1.40,angle=0]{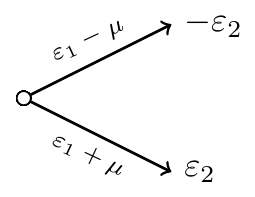}
\label{fig:CmuD}
\end{center}
\end{figure}

\begin{figure}[H]
\begin{center}
\caption{Risk $D$ after omitting the common components of $C(\mu)$ and $D$
}
\vskip 0.4cm
\includegraphics[scale=1.40,angle=0]{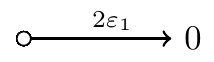}
\label{fig:DD}
\end{center}
\end{figure}

We note that $\mu$ 
occurs as a natural generalization of Pratt's \cite{P64} probability 
premium
to the situation of risk changes with probability mass less than unity.
Indeed, this becomes readily apparent upon omitting the common components of $C(\mu)$ and $D$,
yielding $C(\mu)\setminus \left(C(\mu)\cap D\right)$
and $D\setminus \left(C(\mu)\cap D\right)$ represented in Figures \ref{fig:CmuD} and \ref{fig:DD}, respectively.
\vskip -0.5cm
\noindent
and
\vskip -0.5cm
\noindent When $p_{0}=\varepsilon_{1}=\tfrac{1}{2}$, Pratt's \cite{P64} probability premium occurs as a special case:
it compares a binary risk of obtaining $+\varepsilon_{2}$ and $-\varepsilon_{2}$, initially equally probable,
to the sure status quo.
When $\varepsilon_{1}<\tfrac{1}{2}$ our definition of the probability premium
is a natural generalization of Pratt's definition.\footnote{While the comparison between $C$ and $D$ is a comparison between
two risks (like in Ross \cite{R81}, Machina and Schmeidler \cite{MS89} and Pratt \cite{P90}),
Figures~\ref{fig:CmuD} and \ref{fig:DD} show that it is essentially a Pratt-Arrow-comparison
between a risk and a fixed wealth level with the same probability mass,
which we now allow to be less than unity.}
We show in Section \ref{sec:cra} that our definition of the probability premium satisfies
the appropriate comparative risk aversion results under RDU,
including EU and DT special cases.
Appendix~\ref{sec:nonbinary} further generalizes the definition of the probability premium to apply to non-binary risks.

\setcounter{equation}{0}

\section{First- and Second-Order Probability Aversion}\label{sec:probaversion}

Employing the probability premium thus defined, we state the following definition.

\begin{definition}
A DM is first-order probability averse (seeking) at the point $(w_{0},p_{0})$ if
\begin{equation}
\frac{\partial \mu}{\partial \varepsilon_{1}}\Big|_{\varepsilon_{1}=0^{+}}>(<)\ 0,
\label{eq:1statp}
\end{equation}
and is second-order probability averse (seeking) at $(w_{0},p_{0})$ if
\begin{equation}
\frac{\partial \mu}{\partial \varepsilon_{1}}\Big|_{\varepsilon_{1}=0}=0,\qquad\mathrm{and}\qquad
\frac{\partial \mu^{2}}{\partial \varepsilon_{1}^{2}}\Big|_{\varepsilon_{1}=0^{+}}>(<)\ 0.
\label{eq:2ndatp}
\end{equation}
\end{definition}

In other words, we say that a DM's attitude towards probability is of order 1
if his probability premium $\mu(\varepsilon_{1})$ is not of smaller order than $\varepsilon_{1}$,
i.e., $\mu(\varepsilon_{1})$ is not $o(\varepsilon_{1})$.
Similarly, a DM's attitude towards probability is of order 2
if his probability premium is of smaller order than $\varepsilon_{1}$ but not of smaller order than $\varepsilon_{1}^{2}$,
i.e., $\mu(\varepsilon_{1})$ is $o(\varepsilon_{1})$ but not $o(\varepsilon_{1}^{2})$.

The (strict) signs of the partial derivatives in \eqref{eq:1statp} and \eqref{eq:2ndatp},
determining whether attitude towards probability is of order 1 or of order 2,
have important economic applications;
see Section \ref{sec:risksharing} where these concepts are applied to analyze risk sharing.

We state the following theorem, which is instrumental to our results that follow.
Recall $C$, $D$ and $C(\cdot)$ defined in Figures \ref{fig:C}--\ref{fig:Cmu}.
\begin{theorem}
Let $m>0$.
A second-order probability-averse DM prefers $C(m\varepsilon_{1})$ to $D$
for sufficiently small $\varepsilon_{1}$.
If $m>0$ is small enough, a first-order probability-averse DM prefers $D$ to $C(m\varepsilon_{1})$
for sufficiently small $\varepsilon_{1}$.
\label{th:pa}
\end{theorem}

Theorem \ref{th:pa} reveals that
a first-order probability averter facing risk $C$ may reject to replace it by the actuarially more favorable risk $C(m\varepsilon_{1})$
and opt for the mean-preserving contraction $D$ instead,
even when the probability mass associated with the contraction is very small.

The orders of probability aversion can be viewed as the probability equivalents of the orders of risk aversion.
Indeed, in important work, Segal and Spivak \cite{SS90}
introduce first- and second-order risk aversion.
They consider the risk premium of Pratt \cite{P64} and Arrow \cite{A65,A71}
and analyze its limiting behavior
for a zero-mean risk with 
payoff 
tending to zero.
If the risk premium behaves proportionally to the size of the (squared) payoff of the risk,
the DM exhibits first- (second-)order risk aversion.
Whether attitude towards risk is a first- or second-order phenomenon,
is shown to have important implications for insurance coverage and portfolio choice.
Indeed, a first-order risk-averse DM
may buy full insurance coverage even if the insurance premium carries a positive safety loading,
contrary to a second-order risk-averse DM, who only prefers actuarially fair full insurance coverage.
Likewise, under first-order risk aversion, an investor may choose not to invest in a risky asset despite its mean being positive,
contrary to under second-order risk aversion.

The two measures given by the risk premium and---our generalization of---the probability premium
both evaluate 
how to counteract the effect of a mean-preserving contraction.
That is, both measures capture the intensity of the aversion to mean-preserving spreads.
The orders of probability aversion introduced in this paper
describe how to counteract a mean-preserving contraction marginally
through the probability lens, i.e., by a shift in probabilities,
while the orders of risk aversion describe how a mean-preserving contraction can be off-set marginally
by a change in outcome.

\setcounter{equation}{0}

\section{Risk Sharing}\label{sec:risksharing}

This section presents an application of first- and second-order probability aversion
to a simple risk sharing problem.
It reveals that a first-order probability-averse individual facing a 
risk with a small loss probability
may prefer to 
share his risk in a pool of individuals instead of bearing the risk himself,
even when the other individuals' loss probability 
is 
larger than the
probability of occurrence of the individual's own loss.
By contrast, a second-order probability-averse individual facing this risk only prefers to engage in actuarially fair 
risk sharing.

This result may be viewed as a dual version of Mossin's \cite{M68} Theorem.\footnote{We refer to Segal and Spivak \cite{SS90} and Machina \cite{M13} for an insightful
exposition of Mossin's Theorem,
as well as to the textbook treatment in Eeckhoudt, Gollier and Schlesinger \cite{EGS05}, Chapter 3.}
Indeed, this well-known and seminal result entails that
a second-order risk-averse individual only prefers actuarially fair full insurance,
and opts for a partial insurance contract if the insurance premium carries a positive safety loading.
A first-order risk-averse individual, however, may prefer full insurance
even in the presence of a small positive safety loading.

In Mossin's Theorem the trade-off between risk aversion and a positive safety loading occurs in the ``payoff plane''.
In our result, the trade-off between probability aversion and actuarially unfair loss probabilities occurs in the ``probability plane''.

\subsection{Actuarially Unfavorable Risk Sharing}

Consider an individual who faces a risk $A$ of losing $\ell>0$ with probability $\varepsilon_{1}>0$:
\vskip -0.5cm
\begin{figure}[H]
\begin{center}
\caption{Risk $A$
}
\vskip 0.4cm
\includegraphics[scale=1.40,angle=0]{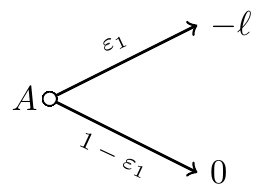}
\label{fig:A}
\end{center}
\end{figure}
\noindent Throughout this section, $\varepsilon_{1}$ is considered to be ``small''.
Of course, any risk-averse individual prefers to replace $A$ by the risk $B$,
which is a mean-preserving contraction of $A$:
\vskip -0.5cm
\begin{figure}[H]
\begin{center}
\caption{Risk $B$
}
\vskip 0.4cm
\includegraphics[scale=1.40,angle=0]{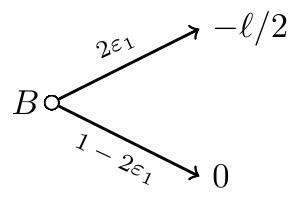}
\label{fig:B}
\end{center}
\end{figure}
\noindent Risk $B$ can be interpreted as the result of a risk sharing contract between two individuals
both facing a mutually exclusive risk of losing $\ell$ with probability $\varepsilon_{1}$.
The two individuals form a pool and share the possible loss---$\ell$ with probability of occurrence $2\varepsilon_{1}$---equally,
each bearing $\ell/2$.

Because $\varepsilon_{1}$ is assumed to be small, the same situation occurs up to the leading order in $\varepsilon_{1}$
when the risks of the two individuals are independent instead of mutually exclusive.
Indeed, under independence, equal risk sharing leads to an allotted risk
\begin{align*}
\left\{
  \begin{array}{lll}
    -\ell, & \mathrm{with\ probability}\ \varepsilon_{1}^{2}&=O(\varepsilon_{1}^{2});\\
    -\ell/2, & \mathrm{with\ probability}\ 2\varepsilon_{1}(1-\varepsilon_{1})&=2\varepsilon_{1}+O(\varepsilon_{1}^{2});\\
    0, & \mathrm{with\ probability}\ (1-\varepsilon_{1})^{2}&=1-2\varepsilon_{1}+O(\varepsilon_{1}^{2}).
  \end{array}
\right.
\end{align*}

More generally, any risk-averse DM prefers to replace $A$ by the risk $B^{(n)}$:
\vskip -0.5cm
\begin{figure}[H]
\begin{center}
\caption{Risk $B^{(n)}$
}
\vskip 0.4cm
\includegraphics[scale=1.40,angle=0]{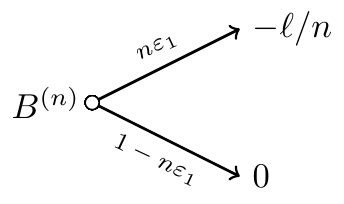}
\label{fig:Bn}
\end{center}
\end{figure}
\noindent where $n\geq 2$.
Indeed, while $B$ is a mean-preserving contraction of $A$, $B^{(n)}$ is a mean-preserving contraction of both $A$ and $B$.
The preference of $B^{(n)}$ to $A$ can be interpreted as a preference for risk sharing among individuals facing risks with the same binary loss distribution,
i.e., actuarially fair risk sharing.

\begin{figure}[H]
\begin{center}
\caption{Risk $A^{\ast}$
}
\vskip 0.4cm
\includegraphics[scale=1.40,angle=0]{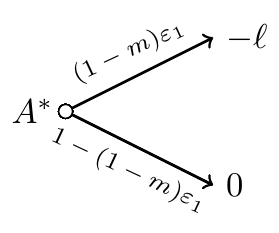}
\label{fig:Aast}
\end{center}
\end{figure}

Theorem \ref{th:pa} now implies that, for sufficiently small $m>0$,
a first-order probability-averse individual
even prefers to replace the risk $A^{\ast}$ represented in Figure \ref{fig:Aast}:
\vskip -0.5cm
\noindent
which is more favorable when compared to $A$,
by the previous risk sharing contract yielding $B^{(n)}$, $n\geq 2$, in Figure \ref{fig:Bn}. 
The preference of $B^{(n)}$ to $A^{\ast}$
can be given the interpretation of a preference for risk sharing among individuals
even when other individuals' losses are slightly more likely to occur
than the individual's own loss.\footnote{Indeed, the probability of a loss in risk $B^{(n)}$, which is $n\varepsilon_{1}$,
is larger than $n$ times the probability of a loss in risk $A^{\ast}$, which is $\left(1-m\right)\varepsilon_{1}$, that is,
$n\varepsilon_{1}>n\left(1-m\right)\varepsilon_{1}$.}

By contrast, a second-order probability-averse individual prefers $A^{\ast}$ to $B$
whenever $m>0$.
In fact, under second-order probability aversion,
$A^{\ast}$ is even preferred to $B^{(n)}$ as soon as $m>0$.
That is, when other individuals' losses are more likely to occur than the individual's own loss,
there is no longer a preference for risk sharing.
A second-order probability-averse individual will only engage in risk sharing when other individuals' losses
have at most the same probability of occurrence
as the individual's own loss.\footnote{Formally, this follows from the fact that under second-order probability aversion
the probability premium that achieves indifference between $A(\mu)$ and $B^{(n)}$
is of order $O(\varepsilon_{1}^{2})$,
while the probability shift from $A$ to $A^{\ast}$ is of order $O(\varepsilon_{1})$.}

Summarizing, we have shown that
under first-order probability aversion a preference for risk sharing exists even when other individuals' losses have slightly larger probability of occurrence;
under second-order probability aversion there is merely a preference for actuarially fair risk sharing.

\subsection{A Graphical Representation}

\begin{figure}[H]
\begin{center}
\caption{The Initial Risk Sharing Problem
}
\vskip 0.37cm
\includegraphics[scale=1.4,angle=0]{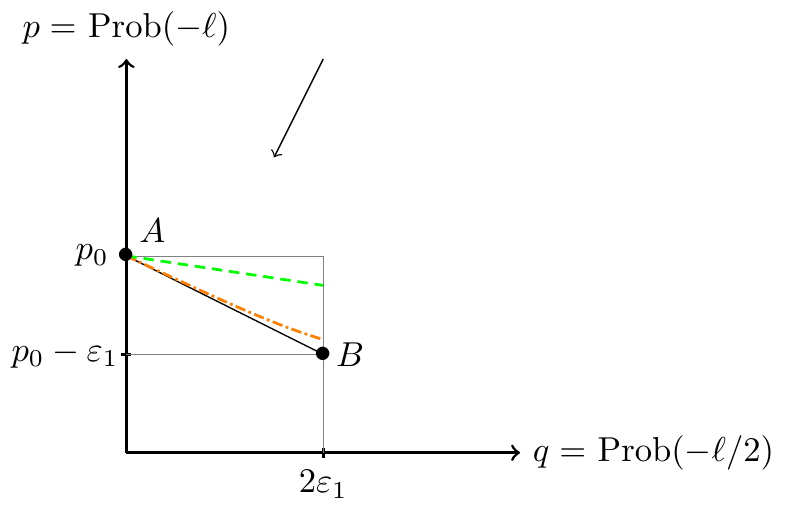}
\label{fig:initial-EL-triangle}
\end{center}
\end{figure}

\begin{figure}[H]
\begin{center}
\caption{The risk corresponding to a point $(q,p)$ in Figure \ref{fig:initial-EL-triangle}
}
\vskip 0.4cm
\includegraphics[scale=1.40,angle=0]{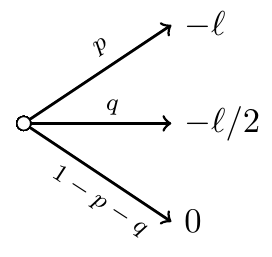}
\label{fig:qp}
\end{center}
\end{figure}

We now provide a graphical representation of the risk sharing problem. 
Figure~\ref{fig:initial-EL-triangle} displays preferences over changes in probabilities
and is an appropriate modification of a Marschak-Machina triangle (\cite{M50}, \cite{M82}),
better suited for our purposes. 
It can be viewed as the dual counterpart of a classic Hirshleifer-Yaari diagram (\cite{H65,H66}, \cite{Y65,Y69})
that displays preferences over changes in outcomes.
\vskip -0.5cm
A point $(q,p)$ in Figure~\ref{fig:initial-EL-triangle} represents a risk
for which outcome $-\ell$ occurs with probability $p$,
outcome $-\ell/2$ occurs with probability $q$,
and 
outcome $0$ occurs with probability $1-p-q$.
That is, the point $(q,p)$ represents the risk given in Figure \ref{fig:qp}:
\vskip -0.5cm
\noindent
Of course, the point $(0,p_{0})$ indicated by $A$ in Figure~\ref{fig:initial-EL-triangle} represents the situation of no risk sharing,
whereas the point $(2\varepsilon_{1},p_{0}-\varepsilon_{1})$ indicated by $B$ represents 
risk sharing.\footnote{The exposition and analysis have so far focused on the situation in which $p_{0}\equiv\varepsilon_{1}$,
hence $p_{0}$ is small just like $\varepsilon_{1}$.
We now consider the more general case in which
we allow for arbitrary $p_{0}$ (with $0<p_{0}<1$ and $0<\varepsilon_{1}\leq \min\{p_{0},1-p_{0}\}$).
That is, $A$ and $B$ in Figure~\ref{fig:initial-EL-triangle} are the analogs of risks $A$ and $B$ in Figures~\ref{fig:A} and~\ref{fig:B},
but $p_{0}$ is no longer restricted to equal $\varepsilon_{1}$.}
The solid chord $AB$ in Figure~\ref{fig:initial-EL-triangle} represents an iso-expected value line. 
It can be viewed as the DM's initial ``budget line'' when sharing risk.

We also illustrate 
in Figure~\ref{fig:initial-EL-triangle}
the indifference curves
of a first-order probability averter (dashed) and a second-order probability averter (dashed-dotted)
that touch the no risk sharing point $(0,p_{0})$.
Such an indifference curve describes the set of points $(q,p)$ that make the DM equally well off as the no risk sharing situation.

\begin{figure}[H]
\begin{center}
\caption{Determining the Indifference Curve
}
\vskip 0.37cm
\includegraphics[scale=1.4,angle=0]{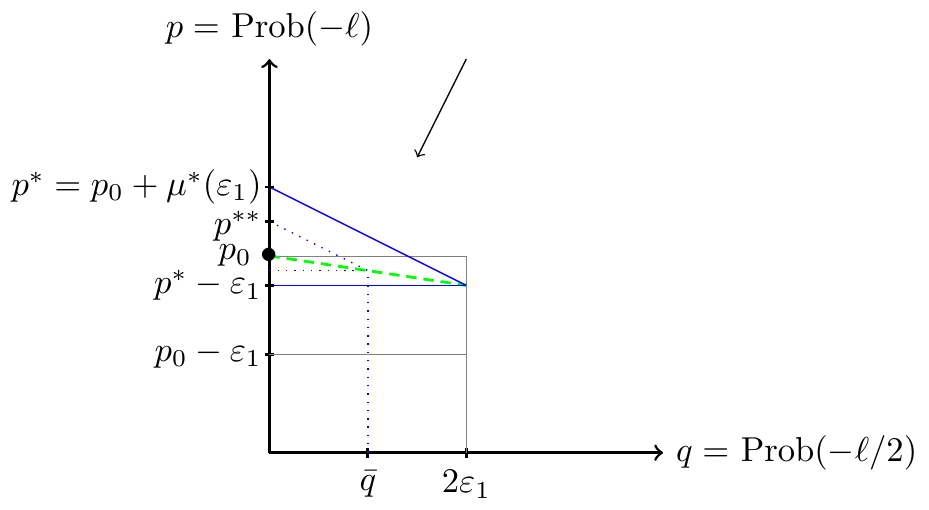}
\label{fig:ind-EL-triangle}
\end{center}
{\footnotesize \textit{Notes:} Starting from the given values of $p_{0}$ and $\varepsilon_{1}$, we determine $p^{\ast}$ and the associated probability premium $\mu^{\ast}(\varepsilon_{1})$
such that $p^{\ast}-\mu^{\ast}(\varepsilon_{1})=p_{0}$, and consider the blue solid triangle.
Note that, by the definition of the probability premium $\mu^{\ast}(\varepsilon_{1})$, the DM is indifferent between the points $(0,p_{0})$ and $(2\varepsilon_{1},p^{\ast}-\varepsilon_{1})$.
Next, for the arbitrarily given value of $\bar{q}$, we determine $p^{\ast\ast}$ and the associated probability premium $\mu^{\ast\ast}\left(\frac{1}{2}\bar{q}\right)$
such that $p^{\ast\ast}-\mu^{\ast\ast}\left(\frac{1}{2}\bar{q}\right)=p^{\ast}-\mu^{\ast}(\varepsilon_{1})=p_{0}$.
The DM is now indifferent between the points $(0,p_{0})$, $(\bar{q},p^{\ast\ast}-\tfrac{1}{2}\bar{q})$ and $(2\varepsilon_{1},p^{\ast}-\varepsilon_{1})$.
See the blue dotted triangle, which is geometrically self-similar to the blue solid triangle.}
\end{figure}

To determine these indifference curves
without imposing additional assumptions on the preference functional,
we proceed as follows, exploiting geometric ``self-similarity''.\footnote{If the functional form of the preference representation were known,
one could easily compute its total differential, by taking the partial derivatives with respect to $q$ and $p$, and equate it to zero.
The slope of the indifference curve would then be given by minus the ratio of the representation's partial derivatives with respect to $q$ and $p$.}
For given $p_{0}$, $\varepsilon_{1}$ and arbitrary $\bar{q}$, with $0<\bar{q}<2\varepsilon_{1}$,
we find $p^{\ast\ast}$, with $0<p^{\ast\ast}<p^{\ast}:=p_{0}+\mu^{\ast}(\varepsilon_{1})$, such that
\begin{equation*}
p^{\ast}-\mu^{\ast}(\varepsilon_{1})=p^{\ast\ast}-\mu^{\ast\ast}\left(\frac{1}{2}\bar{q}\right),
\end{equation*}
hence we obtain
\begin{equation*}
p^{\ast\ast}=p^{\ast}-\mu^{\ast}(\varepsilon_{1})+\mu^{\ast\ast}\left(\frac{1}{2}\bar{q}\right).
\end{equation*}
The indifference curve is then given by
\begin{align}
p(\bar{q})&=p^{\ast\ast}-\frac{1}{2}\bar{q}\nonumber\\
&=p^{\ast}-\mu^{\ast}(\varepsilon_{1})+\mu^{\ast\ast}\left(\frac{1}{2}\bar{q}\right)-\frac{1}{2}\bar{q}.
\label{eq:indiff}
\end{align}
See Figure~\ref{fig:ind-EL-triangle}, where this procedure of constructing the indifference curve by geometric self-similarity is illustrated for the first-order risk averter.
(The analogous figure for the second-order risk averter is similar.)
\vskip -0.5cm
\noindent

From Eqn.~\eqref{eq:indiff}, one readily verifies that the slope of the indifference curve in $A$, i.e., at $(0,p_{0})$,
is given by
\begin{equation*}
\frac{\partial p(\bar{q})}{\partial \bar{q}}\Big|_{\bar{q}=0^{+}}
=-\frac{1}{2}+\frac{1}{2}\frac{\partial \mu^{\ast\ast}\left(\tfrac{1}{2}\bar{q}\right)}{\partial \bar{q}}\Big|_{\bar{q}=0^{+}}.
\end{equation*}
Hence, when the DM exhibits attitude towards probability of order 2,
the indifference curve is tangent to the iso-expected value line at $(0,p_{0})$;
see the dashed-dotted curve in Figure~\ref{fig:initial-EL-triangle}.
However, when the DM's attitude towards probability is of order 1,
the indifference curve at $(0,p_{0})$ is not tangent to, but departs faster from, the iso-expected value line;
see the dashed curve in Figure~\ref{fig:initial-EL-triangle}.

\begin{figure}[H]
\begin{center}
\caption{Actuarially Favorable Risk Sharing
}
\vskip 0.37cm
\includegraphics[scale=1.4,angle=0]{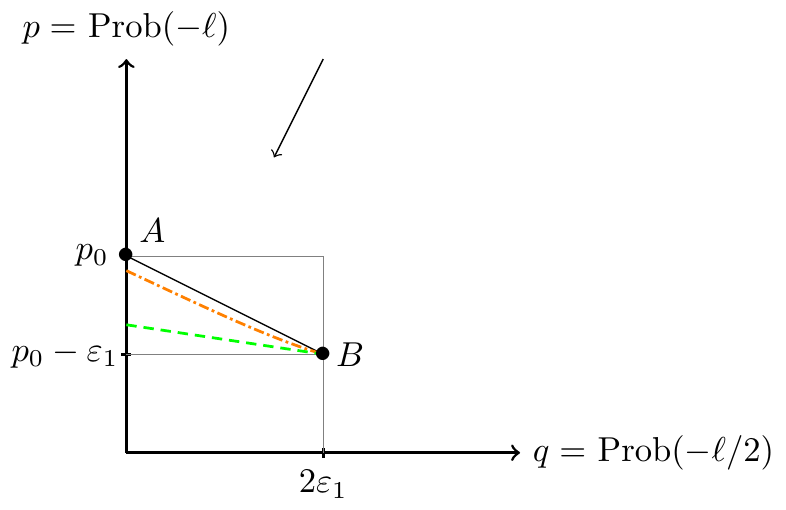}
\label{fig:initial-optimum-EL-triangle}
\end{center}
\end{figure}

The DM seeks to move the indifference curve southwestward (as indicated by the arrow),
towards the origin where he is better off, but is subject to the budget restriction.
Therefore, when the choice menu consists of $A$ and $B$, both the first- and second-order risk averter prefer $B$;
see Figure~\ref{fig:initial-optimum-EL-triangle}.
\vskip -0.5cm

\begin{figure}[H]
\begin{center}
\caption{Actuarially \textit{Un}favorable Risk Sharing
}
\vskip 0.37cm
\includegraphics[scale=1.4,angle=0]{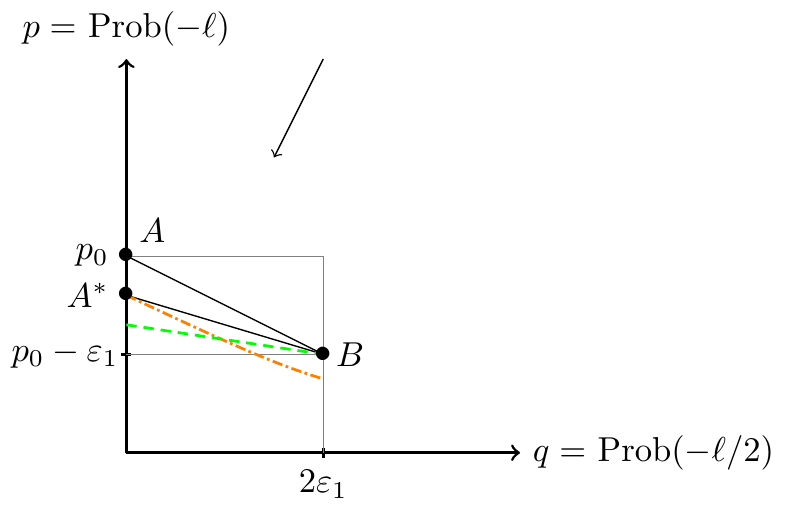}
\label{fig:EL-triangle}
\end{center}
\end{figure}

Now imagine a change in the budget line from $AB$ to $A^{\ast}B$; see Figure~\ref{fig:EL-triangle}.
Of course, the risk $A^{\ast}$ is more favorable than the risk $A$.
In this new situation, risk $A^{\ast}$ corresponds to no risk sharing
whereas risk $B$ corresponds to risk sharing with an individual who has a larger loss probability.
Figure~\ref{fig:EL-triangle} now reveals that a first-order probability averter may still prefer the risk sharing contract $(2\varepsilon_{1},p_{0}-\varepsilon_{1})$
as long as $A^{\ast}$ does not become too favorable.
\vskip -0.5cm
\noindent Indeed, after moving his indifference curve (dashed) southwestward as far as possible in the new situation,
it lies below the new budget line, except at the point $(2\varepsilon_{1},p_{0}-\varepsilon_{1})$.
By contrast, the second-order probability averter will always opt to bear the risk himself
instead of engaging in actuarially unfavorable risk sharing. 
Indeed, 
after moving his indifference curve southwestward towards the origin,
his indifference curve (dashed-dotted) lies below the new budget line
except in $A^{\ast}$.

Thus, even when others have more unfavorable loss probabilities than the individual himself,
first-order probability aversion may induce a preference for 
risk sharing when facing a risk with a small loss probability.
By contrast, a second-order probability averter facing the same risk
does not engage in actuarially unfavorable risk sharing.

\subsection{Risk Sharing Conclusion}

Of course, when the probability mass associated with the unfavorable outcome of the risk to be shared is sufficiently large,
and the discrepancy between the individual's own loss probability and that of the other individual(s) is not too large,
both a first- and a second-order risk averter benefit from actuarially unfavorable risk sharing.
The analysis in the section therefore quite naturally restricts attention to the case of small loss probabilities,
which are also practically most relevant.

Our analysis reveals that in this setting a first-order probability averter may immediately start to share his risk whenever the opportunity arises,
even when the individual's own loss probability is smaller than that of the other individual(s)
(as long as the discrepancy between the individual's own loss probability and that of the other individual(s) does not become too large).
By contrast, a second-order probability averter only engages in actuarially fair risk sharing.

\setcounter{equation}{0}

\section{Rank-Dependent Utility}\label{sec:rdu}

In this section, we analyze the probability premium and attitude towards probability
under Yaari's \cite{Y87} dual theory of choice under risk (DT) and,
more generally and behaviorally more relevant, under the rank-dependent utility model (RDU; Quiggin \cite{Q82}).

Under DT,
a risk with cumulative distribution function $F$ is evaluated according to
\begin{equation}
\int x \,\mathrm{d}h\left(F(x)\right),
\label{eq:DT}
\end{equation}
where, here and in the sequel, the integral runs over the support of $F$.\footnote{Formally, \eqref{eq:DT} is a Riemann-Stieltjes integral.}
The probability weighting (distortion) function $h$ maps the unit interval onto itself,
satisfies $h(0)=0$ and $h(1)=1$, and is supposed to be non-decreasing.
For $n$-state risks with payoffs $x_{1}\leq\cdots\leq x_{n}$
and associated probabilities $p_{1},\ldots,p_{n}$,
\eqref{eq:DT} can be expressed as
\begin{equation*}
\sum_{i=1}^{n}\left(h\left(\sum_{j=1}^{i}p_{j}\right)-h\left(\sum_{j=1}^{i-1}p_{j}\right)\right)x_{i},
\end{equation*}
with $\sum_{j=1}^{0}p_{j}=0$ by convention.

The DT evaluation occurs as a special case of the RDU model
which evaluates a risk with cumulative distribution function $F$ by the preference functional
\begin{equation}
\int U(x) \,\mathrm{d}h\left(F(x)\right).
\label{eq:RDU}
\end{equation}
The utility function $U$ is unique up to positive affine transformations
and is supposed to be non-decreasing,
and $h$ is as above.\footnote{For $n$-state risks with payoffs $x_{1}\leq\cdots\leq x_{n}$
and associated probabilities $p_{1},\ldots,p_{n}$,
\eqref{eq:RDU} simplifies to
\begin{equation*}
\sum_{i=1}^{n}\left(h\left(\sum_{j=1}^{i}p_{j}\right)-h\left(\sum_{j=1}^{i-1}p_{j}\right)\right)U(x_{i}).
\end{equation*}}
Henceforth, unless otherwise stated, we assume $U$ and $h$ to be twice continuously differentiable
with $U'>0$ and $h'>0$.
Of course, the EU model occurs as a special case when $h$ is the identity,
while the DT model arises when $U$ is affine.\footnote{Equivalently,
the RDU evaluation occurs by distorting \textit{de}cumulative probabilities $S(\cdot)=1-F(\cdot)$
following Yaari \cite{Y87}
rather than cumulative probabilities, as follows:
\begin{align*}
\int_{a}^{b} U(x) \,\mathrm{d}h\left(F(x)\right)
=\int_{a}^{b} U(x) \,\mathrm{d}\left(1-\bar{h}\left(S(x)\right)\right)
=\int_{a}^{b} \bar{h}\left(S(x)\right)\,\mathrm{d}U(x) ,
\end{align*}
where $\bar{h}(p)=1-h(1-p)$ and we normalize $U(a)=0$.
Then, for $n$-state risks, we have the evaluation
\begin{equation*}
\sum_{i=1}^{n}\left(\bar{h}\left(1-\sum_{j=1}^{i-1}p_{j}\right)-\bar{h}\left(1-\sum_{j=1}^{i}p_{j}\right)\right)U(x_{i}).
\end{equation*}
Note that $\bar{h}(0)=0$, $\bar{h}(1)=1$,
that $\bar{h}$ is non-decreasing, 
and that $\bar{h}''>0$ is equivalent to $h''<0$.
The condition $h''<0$, jointly with $U''<0$,
means that the RDU DM is ``strongly risk averse'' in the sense of aversion to any mean-preserving increase in risk \`a la Rothschild and Stiglitz \cite{RS70}
(Chew, Karni and Safra \cite{CKS87} and Ro\"ell \cite{R87}).}

\subsection{Local Approximation to the Probability Premium}\label{sec:localappRDU}

Under DT, the probability premium $\mu$
defined in Section \ref{sec:probpremium}, Figure \ref{fig:Cmu},
occurs as the solution to\footnote{The terms associated
with the low and high wealth level states 
partially cancel.
Those terms that cancel are suppressed.}
\begin{align*}
&\left(h(p_{0}+\varepsilon_{1})-h(p_{0}-\varepsilon_{1})\right)w_{0}\nonumber\\
&\qquad=\left(h(p_{0}-\mu)-h(p_{0}-\varepsilon_{1})\right)\left(w_{0}-1\right)
+\left(h(p_{0}+\varepsilon_{1})-h(p_{0}-\mu)\right)\left(w_{0}+1\right),
\end{align*}
which reduces to
\begin{equation}
0=h(p_{0}-\varepsilon_{1})-2h(p_{0}-\mu)+h(p_{0}+\varepsilon_{1}).
\label{eq:ppDT}
\end{equation}

Eqn.~\eqref{eq:ppDT} does not in general admit an explicit solution.
As in the local risk aversion approach of Pratt \cite{P64} and Arrow \cite{A65,A71},
a very insightful dissection of the probability premium
can be obtained by considering ``small risks''.
The resulting local approximation
disentangles the complex interplay between the probability distribution of the risk change
and the DM's preferences that the probability premium depends on.
Throughout, our designation ``small'' in ``small risks'' will refer to risks that have
small (unconditional) probabilities under DT
and both small payoffs (i.e., outcomes)
and small (unconditional) probabilities under RDU.

Indeed, from~\eqref{eq:ppDT}, a first-order Taylor series expansion of $h(p_{0}-\mu)$ around $p_{0}$,
and a second-order Taylor series expansion of $h(p_{0}\pm\varepsilon_{1})$ around $p_{0}$,
yield
\begin{equation*}
0\simeq\varepsilon_{1}^{2} h''(p_{0})+2\mu h'(p_{0}),
\end{equation*}
hence
\begin{equation}
\mu\simeq -\frac{1}{2}\varepsilon_{1}^{2}\frac{h''(p_{0})}{h'(p_{0})}.
\label{eq:ppDTapp}
\end{equation}

Next, we consider the probability premium under RDU,
obtained as the solution to
\begin{align}
&\left(h(p_{0}+\varepsilon_{1})-h(p_{0}-\varepsilon_{1})\right)U(w_{0})\nonumber\\
&\qquad=\left(h(p_{0}-\mu)-h(p_{0}-\varepsilon_{1})\right)U(w_{0}-\varepsilon_{2})
+\left(h(p_{0}+\varepsilon_{1})-h(p_{0}-\mu)\right)U(w_{0}+\varepsilon_{2}).
\label{eq:ppRDU}
\end{align}
Just like under DT, we approximate the solution to (\ref{eq:ppRDU}) by suitable Taylor series expansions.
We first invoke a second-order Taylor series expansion of $U(w_{0}\pm \varepsilon_{2})$ around $w_{0}$
to obtain
\begin{align*}
2h(p_{0}-\mu)\simeq h(p_{0}-\varepsilon_{1})+h(p_{0}+\varepsilon_{1})+\left(h(p_{0}+\varepsilon_{1})-h(p_{0}-\varepsilon_{1})\right)\frac{1}{2}\varepsilon_{1}\frac{U''(w_{0})}{U'(w_{0})}.
\end{align*}
Next, a first-order Taylor series expansion of $h(p_{0}-\mu)$ around $p_{0}$,
and a second-order Taylor series expansion of $h(p_{0}\pm\varepsilon_{1})$ around $p_{0}$,
yields
\begin{equation}
\mu\simeq-\frac{1}{2}\varepsilon_{1}\varepsilon_{2}\frac{U''(w_{0})}{U'(w_{0})}-\frac{1}{2}\varepsilon_{1}^{2}\frac{h''(p_{0})}{h'(p_{0})}.
\label{eq:ppRDUapp}
\end{equation}
Eqn.~\eqref{eq:ppRDUapp} reveals that the interplay between the ratios of second and first derivatives
of the utility function and probability weighting function
are central in dictating the local approximation to the probability premium.
As we will see in Section \ref{sec:cra}, the
local indexes $\tfrac{U''(w_{0})}{U'(w_{0})}$ and $\tfrac{h''(p_{0})}{h'(p_{0})}$
are the appropriate measures of the intensity of risk aversion under RDU
also from a global perspective.

\subsection{Maxiance and Variance}\label{sec:dualmom}

To fully appreciate the local approximations \eqref{eq:ppDTapp} and \eqref{eq:ppRDUapp} to the probability premium
under the DT and RDU models,
we introduce the \textit{maxiance} or second dual moment about the mean.

Of course, for a risk $\tilde{\varepsilon}$ with cumulative distribution function $F$,
the first (primal and dual) moment can be expressed as
\begin{equation}
\texttt{m}:=\mathbb{E}\left[\tilde{\varepsilon}\right]=\int x\,\mathrm{d}F(x),
\end{equation}
whereas the second primal moment about the mean, or variance, is given by
\begin{equation}
\texttt{m}_{2}:=\mathbb{E}\left[\left(\tilde{\varepsilon}-\texttt{m}\right)^{2}\right]=\int \left(x-\texttt{m}\right)^{2}\,\mathrm{d}F(x).
\label{eq:var}
\end{equation}
The second dual moment about the mean, or maxiance, is defined by
\begin{equation}
\bar{\texttt{m}}_{2}:=\mathbb{E}\left[\max\left(\tilde{\varepsilon}^{(1)},\tilde{\varepsilon}^{(2)}\right)\right]-\mathbb{E}\left[\tilde{\varepsilon}\right],
\end{equation}
where $\tilde{\varepsilon}^{(1)}$ and $\tilde{\varepsilon}^{(2)}$
are two independent copies of $\tilde{\varepsilon}$.

The maxiance represents the expectation of the largest
among two independent draws of the risk $\tilde{\varepsilon}$
beyond the mean.
It can be expressed as
\begin{equation}
\bar{\texttt{m}}_{2}=\int \left(x-\texttt{m}\right)\,\mathrm{d}\left(F(x)\right)^{2}.
\label{eq:max}
\end{equation}
Comparing \eqref{eq:var} and \eqref{eq:max} shows that the maxiance is a measure of risk in the ``probability plane'',
contrary to the variance, which is a measure of risk in the ``payoff plane''.

For a zero-mean risk, one may verify that
\begin{equation}
\mathbb{E}\left[\min\left(\tilde{\varepsilon}^{(1)},\tilde{\varepsilon}^{(2)}\right)\right]=-\bar{\texttt{m}}_{2}.
\label{eq:min}
\end{equation}
Arguably the \textit{miniance} in \eqref{eq:min},
i.e., the worst among two independent draws of $\tilde{\varepsilon}$,
is a more natural measure of ``risk'',
but from \eqref{eq:min} we observe that for zero-mean risks
the miniance coincides with the maxiance up to a change of sign.

Dual moments play the same role in inverse (or dual) stochastic dominance as primal moments in stochastic dominance;
see Muliere and Scarsini \cite{MS89} and Eeckhoudt, Laeven and Schlesinger \cite{ELS20}.
We also note that the first and second primal moments are generated by linear and quadratic utility functions in EU,
while the first and second dual moments correspond to linear and quadratic probability weighting functions in DT.
We refer to Eeckhoudt and Laeven \cite{EL20} and the references therein for further details about the maxiance.

\begin{figure}[H]
\begin{center}
\caption{$D$ to $C$: Mean-Preserving Spread.
}
\vskip 0.4cm
\includegraphics[scale=1.40,angle=0]{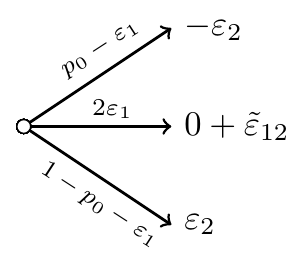}
\label{fig:DtoC}
\end{center}
\end{figure}

Employing the variance and the maxiance
we can re-write \eqref{eq:ppRDUapp} as follows:
\begin{equation}
\mu\simeq\frac{\texttt{m}_{2}}{2\texttt{Py}}\left(-\frac{U''(w_{0})}{U'(w_{0})}\right)
+\frac{\bar{\texttt{m}}_{2}}{2\texttt{Py}}\left(-\frac{h''(p_{0})}{h'(p_{0})}\right),
\label{eq:ppRDUappmax}
\end{equation}
where $\texttt{m}_{2}$ and $\bar{\texttt{m}}_{2}$ are the unconditional variance and maxiance of the risk $\tilde{\varepsilon}_{12}$
describing the mean-preserving spread from $D$ to $C$.
Unconditionally, the risk $\tilde{\varepsilon}_{12}$ assumes the values $\pm\varepsilon_{2}$ each with probability $\varepsilon_{1}$;
see Figure \ref{fig:DtoC}
which shows that $C$ is obtained from $D$ by attaching the binary zero-mean risk $\tilde{\varepsilon}_{12}$
taking the values $\pm\varepsilon_{2}$ to the intermediate branch of $D$.
\vskip 0.4cm
\noindent Furthermore, $\texttt{Py}$ denotes the $1$-norm (or absolute-value norm or $L_{1}$-norm)
of the payoffs of $\tilde{\varepsilon}_{12}$.\footnote{One readily verifies that
for $\tilde{\varepsilon}_{12}$, unconditionally,
$\texttt{m}_{2}=2\varepsilon_{1}\varepsilon_{2}^{2}$, $\bar{\texttt{m}}_{2}=2\varepsilon_{1}^{2}\varepsilon_{2}$,
and $\texttt{Py}=2\varepsilon_{2}$.}

Note that, when $\varepsilon_{1}=\tfrac{1}{2}$, the first term in \eqref{eq:ppRDUappmax}
agrees with Pratt's \cite{P64} local approximation to the probability premium in the EU model.\footnote{We note that in Pratt \cite{P64},
the probability premium is originally defined as the \textit{ex post} difference between the probability of the favorable outcome
and the probability of the unfavorable outcome, i.e., $2\mu$.
Pratt's convention is not followed in the recent literature,
where the probability premium is the \textit{ex post} shift in the probability from the unfavorable outcome
to the favorable outcome, i.e., $\mu$;
see e.g., Liu and Neilson \cite{LN19}.}
Furthermore, the second term in \eqref{eq:ppRDUappmax} can be recognized as the local approximation to the probability premium
in the DT model; see \eqref{eq:ppDTapp}.
Thus, the local approximation to the probability premium under the RDU model occurs as the (suitably scaled) sum of the EU and DT probability premia,
where the variance and maxiance are on equal footing.

In Appendix \ref{sec:nonbinary}, we show that the local approximation to the RDU probability premium in \eqref{eq:ppRDUappmax}
also generalizes naturally to non-binary risks.
Appendix \ref{sec:nonbinary} focuses on the RDU model of which both the EU and DT models are special cases.
We note that both the local approximation in \eqref{eq:ppRDUappmax} and the results in Appendix \ref{sec:nonbinary}
are novel already when restricted to the EU model.
Indeed, Pratt \cite{P64} only considers the EU model with $p_{0}=\varepsilon_{1}=\tfrac{1}{2}$.

\subsection{Attitude Towards Probability}

Under the standing assumption that $U$ and $h$ are twice continuously differentiable
with $U'>0$ and $h'>0$,
we state the following theorem.
\begin{theorem}
A DT DM
has attitude towards probability of second order
at the points where $h''\neq 0$.
An EU and RDU DM
has attitude towards probability of first order
at the points where $U''\neq 0$
(for almost every $\varepsilon_{2}$).
\label{th:ordpa}
\end{theorem}

Now consider the DT model and suppose that the probability weighting function $h:[0,1]\rightarrow[0,1]$
is non-decreasing and not differentiable at $p_{0}$,
but has left and right derivatives $h'_{-}(p_{0})$ and $h'_{+}(p_{0})$,
with $h'_{-}(p_{0})\neq h'_{+}(p_{0})$.
Then we state the following result.
\begin{theorem}
A DT DM
has attitude towards probability of first order
at the points where the probability weighting function is not differentiable
but has left and right derivatives.
\label{th:ordpanondiff}
\end{theorem}

An example of such a probability weighting function is given by
\begin{align}
h(p)=\left\{
       \begin{array}{ll}
         \frac{p}{1-p_{0}}, & 0\leq p\leq 1-p_{0};\\
         1, & 1-p_{0}<p\leq 1.
       \end{array}
     \right.
\end{align}
In financial risk management, this probability weighting function gives rise to that Average-Value-at-Risk (AV@R).

We thus find that RDU and EU maximizers usually exhibit first-order probability aversion,
while DT DMs with smooth probability weighting function are second-order probability averters.
From Segal and Spivak \cite{SS90} we already know that RDU and DT maximizers usually exhibit first-order risk aversion,
while EU DMs with smooth utility function are risk averse of order 2.
That is, under the RDU model,
both probability aversion and risk aversion are usually first-order phenomena.

We note also that expected value maximization---popular in operations research (e.g., in optimal stopping)---,
law-invariant comonotonic coherent risk measures---popular in financial risk management (e.g., F\"ollmer and Schied \cite{FS11}, Chapter 4)---,
and linear (or affine) utility models---popular in economics especially for small stakes---,
usually induce second-order probability aversion.

\subsection{Comparative Risk Aversion}\label{sec:cra}

In this subsection, we show that the local approximation to the RDU probability premium can be naturally linked
to global results on comparative risk aversion under RDU.
Note first that \eqref{eq:ppRDUappmax} reveals that, for small zero-mean risks,
the probability premium $\mu$ is affine in the primal and dual local indexes of absolute risk aversion
\begin{equation*}
\left(-\frac{U''(w_{0})}{U'(w_{0})}\right)\qquad\mathrm{and}\qquad\left(-\frac{h''(p_{0})}{h'(p_{0})}\right).
\end{equation*}
Next, note that the definition of the probability premium in Section \ref{sec:probpremium}, Figure \ref{fig:Cmu},
applies irrespective of whether $\varepsilon_{1}$ and $\varepsilon_{2}$ are ``small'' or ``large''.\footnote{Only
$0<\varepsilon_{1}\leq \min\{p_{0},1-p_{0}\}$ and $\varepsilon_{2}>0$ are required.}

\begin{theorem}
Consider two RDU maximizers with utility function $U_{i}$ and probability weighting function $h_{i}$, $i=1,2$.
Let $\mu_{i}\left(p_{0},w_{0},\varepsilon_{1},\varepsilon_{2}\right)$ be their respective probability premia,
which occur as the solutions to \eqref{eq:ppRDU}.
Then the following statements are equivalent.
\begin{itemize}
\item[(i)] $-\frac{U_{2}''(x)}{U_{2}'(x)}\geq -\frac{U_{1}''(x)}{U_{1}'(x)}$ and $-\frac{h_{2}''(p)}{h_{2}'(p)}\geq -\frac{h_{1}''(p)}{h_{1}'(p)}$
$\quad$ for all $x\in\mathbb{R}$ and all $p\in(0,1)$.
\item[(ii)] $\mu_{2}(p_{0},w_{0},\varepsilon_{1},\varepsilon_{2})\geq \mu_{1}(p_{0},w_{0},\varepsilon_{1},\varepsilon_{2})$
$\quad$ for all $w_{0}\in\mathbb{R}$, all $\varepsilon_{2}>0$, and all $0<\varepsilon_{1}\leq\{p_{0},1-p_{0}\}<1$.
\end{itemize}
\label{th:cra}
\end{theorem}
The statements apply \textit{mutatis mutandis} when the domain of the utility functions is only a subset of $\mathbb{R}$.

From Pratt \cite{P64}, Yaari \cite{Y86} and Eeckhoudt and Laeven \cite{EL20}, statement~(i) is known to be equivalent to
$U_{2}$ and $h_{2}$ being \textit{concavifications}
of $U_{1}$ and $h_{1}$, i.e.,
there exist functions $\phi$ and $\psi$ with
$\phi'>0$, $\phi''\leq 0$, $\psi'>0$, $\psi''\leq 0$, $\psi(0)=0$ and $\psi(1)=1$
such that
\begin{equation*}
U_{2}(x):=\phi\left(U_{1}(x)\right)\qquad\mathrm{and}\qquad h_{2}(p):=\psi\left(h_{1}(p)\right),
\end{equation*}
for all $x\in\mathbb{R}$ and all $p\in[0,1]$.

Theorem~\ref{th:cra} generalizes Pratt's results on the probability premium
in two directions:
to risk changes with probability mass less than unity
and, more significantly, to RDU.
Furthermore, as the DT model occurs as a special case of the RDU model,
it is also covered by the results in Theorem~\ref{th:cra}.
The generality of the results in Theorem~\ref{th:cra},
and in particular the joint presence of both the utility function and the probability weighting function,
makes its proof more complex than that in Pratt \cite{P64}.
The proof of Theorem~\ref{th:cra} is in Appendix~\ref{sec:proofs}
and relies on analyzing the total differential of the RDU functional.

Global results on comparative risk aversion under RDU are already in
Yaari \cite{Y86}, Chew, Karni and Safra \cite{CKS87}, and Ro\"ell \cite{R87}.
As $D$ is a mean-preserving contraction of $C$,
these existing global results may in principle be invoked to imply (i)$\Rightarrow$(ii),
although none of the existing results under RDU refers to the probability premium.
The implication (ii)$\Rightarrow$(i) is new.
It establishes the formal connection between a local approach to the probability premium
and global risk aversion within RDU.
An analog of Theorem~\ref{th:cra} for the risk premium is in Eeckhoudt and Laeven \cite{EL20}.

\subsection{Comparison to the Risk Premium}

\begin{figure}[H]
\begin{center}
\caption{Risk $D(\lambda)$
}
\vskip 0.4cm
\includegraphics[scale=1.40,angle=0]{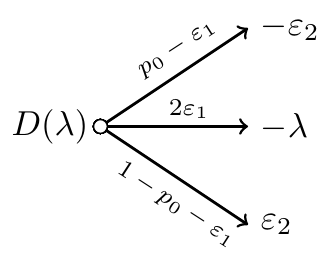}
\label{fig:Dlambda}
\end{center}
\end{figure}

As an alternative to the construction of the probability premium in Section~\ref{sec:probpremium}, Figure~\ref{fig:Cmu},
we could have adapted the outcome of the intermediate state of $D$ from 0 to $-\lambda$
such that the DM becomes indifferent between $C$ and $D(\lambda)$; see Figure~\ref{fig:Dlambda}.
\vskip -0.5cm
\noindent (Of course, $D(0)\equiv D$.)
This would naturally gives rise to the risk premium $\lambda$ associated to the risk change from $C$ to $D$;
see Eeckhoudt and Laeven \cite{EL20}.
One easily verifies that the risk premium of Pratt \cite{P64} and Arrow \cite{A65,A71} occurs as a special case
in the comparison between $C$ and $D(\lambda)$ when $p_{0}=\varepsilon_{1}=\tfrac{1}{2}$.
Then, one compares a risk to a sure loss equal to the risk premium. 

Eeckhoudt and Laeven \cite{EL20} develop a local approximation to the risk premium $\lambda$.
Their approximation takes the form
\begin{align}
\lambda &\simeq -\frac{1}{2}\varepsilon_{2}^{2}\frac{U''(w_{0})}{U'(w_{0})}-\frac{1}{2}\varepsilon_{1}\varepsilon_{2}\frac{h''(p_{0})}{h'(p_{0})}\nonumber\\
&=\frac{\texttt{m}_{2}}{2\texttt{Pr}}\left(-\frac{U''(w_{0})}{U'(w_{0})}\right)
+\frac{\bar{\texttt{m}}_{2}}{2\texttt{Pr}}\left(-\frac{h''(p_{0})}{h'(p_{0})}\right),
\label{eq:rpRDUappmax}
\end{align}
where $\texttt{Pr}$ is the total unconditional probability mass associated to $\tilde{\varepsilon}_{12}$.\footnote{Observe that,
for $\tilde{\varepsilon}_{12}$, $\texttt{Pr}=2\varepsilon_{1}$.}
Just like in \eqref{eq:ppRDUappmax}, the local approximations to the EU and DT risk premia
follow as special cases by setting $h(p)\equiv p$ for EU and $U(x)\equiv x$ for DT.

Upon comparing \eqref{eq:ppRDUappmax} and \eqref{eq:rpRDUappmax},
we obtain the following link between the risk and probability premia:
\begin{equation}
\mu\texttt{Py} = \lambda\texttt{Pr}.
\label{eq:rppplink}
\end{equation}
Observe that the units of measurement on both sides of \eqref{eq:rppplink} agree,
as $\mu$ and $\texttt{Pr}$ are dimensionless.

\setcounter{equation}{0}

\section{Conclusion}\label{sec:Con}

For many years, and despite its appeal, the probability premium of Pratt \cite{P64}
has been overshadowed by its more prevalent siblings
given by the risk premium and the local index of absolute risk aversion.

In this paper, we have generalized Pratt's construction of the probability premium
and have used it to introduce and analyze attitudes towards probability.
By developing a local approximation to the probability premium under rank-dependent utility
as a by-product of significant interest in its own right,
we have shown that rank-dependent and expected utility maximizers usually display
attitude towards probability of order 1,
whereas dual theory maximizers---as well as expected value maximizers,
investors using law-invariant comonotonic coherent risk measures,
and economic agents using affine utility models---typically exhibit attitude towards probability of order 2.

We have illustrated the relevance of first- and second-order probability aversion
in a problem of risk sharing.
Our results reveal that for a first-order risk averter facing a risk with a small loss probability
there are immediately significant benefits to risk sharing opportunities.
He may therefore opt for risk sharing even when actuarially unfavorable.
By contrast, for a second-order risk averter the initial benefits of risk sharing opportunities are negligible.
A second-order risk averter 
likes actuarially fair risk sharing.
However, when facing a risk with a small loss probability, he always avoids actuarially unfavorable risk sharing.
That is,
only first-order probability averters like to 
engage in risk sharing
if the probability of the occurrence of other individuals' losses is 
more unfavorable
than that of the individual's own loss,
when faced with a risk with a small loss probability. 

We hope that our approach and results contribute to 
a 
widespread analysis and use of the appealing concept of the probability premium.

\newpage

\begin{appendices}

\setcounter{equation}{0}

\section{Proofs}\label{sec:proofs}

\noindent {\sc Proof of Theorem~\ref{th:pa}.}
Consider the probability premium $\mu(\varepsilon_{1})$,
which makes the DM indifferent in the comparison between $C(\mu)$ (Figure~\ref{fig:Cmu}) and $D$ (Figure~\ref{fig:D}).
In addition, consider the probability premium $\mu(\varepsilon_{1},m)$,
obtained as the solution that makes the DM indifferent in the comparison between the risk $C(\mu,m)$:
\vskip -0.5cm
\begin{figure}[H]
\begin{center}
\caption{Risk $C(\mu,m)$
}
\vskip 0.4cm
\includegraphics[scale=1.40,angle=0]{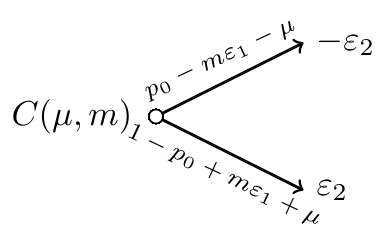}
\label{fig:Cmmu}
\end{center}
\end{figure}
\noindent and $D$ (Figure~\ref{fig:D}).
Of course,
\begin{equation*}
\mu(\varepsilon_{1})=m\varepsilon_{1}+\mu(\varepsilon_{1},m).
\end{equation*}
Hence,
\begin{equation*}
\frac{\partial \mu(\varepsilon_{1})}{\partial \varepsilon_{1}}=m+\frac{\partial \mu(\varepsilon_{1},m)}{\partial \varepsilon_{1}}.
\end{equation*}
Furthermore,
\begin{equation*}
\lim_{\varepsilon_{1}\rightarrow 0^{+}}\frac{\partial \mu(\varepsilon_{1})}{\partial \varepsilon_{1}}
=m+\lim_{\varepsilon_{1}\rightarrow 0^{+}}\frac{\partial \mu(\varepsilon_{1},m)}{\partial \varepsilon_{1}}.
\end{equation*}
If the DM is first-order probability averse,
then
\begin{equation*}
\lim_{\varepsilon_{1}\rightarrow 0^{+}}\frac{\partial \mu(\varepsilon_{1})}{\partial \varepsilon_{1}}>0.
\end{equation*}
Thus, for
\begin{equation*}
m<\frac{\partial \mu(\varepsilon_{1})}{\partial \varepsilon_{1}}\Big|_{\varepsilon_{1}=0^{+}}
\end{equation*}
we have
\begin{equation*}
\frac{\partial \mu(\varepsilon_{1},m)}{\partial \varepsilon_{1}}\Big|_{\varepsilon_{1}=0^{+}}>0.
\end{equation*}
If the DM is second-order probability averse,
then
\begin{equation*}
\lim_{\varepsilon_{1}\rightarrow 0^{+}}\frac{\partial \mu(\varepsilon_{1})}{\partial \varepsilon_{1}}=0
\qquad\mathrm{and}\qquad
\lim_{\varepsilon_{1}\rightarrow 0^{+}}\frac{\partial \mu(\varepsilon_{1},m)}{\partial \varepsilon_{1}}=-m<0,
\end{equation*}
which proves the stated result.
\qed

\noindent {\sc Proof of Theorem~\ref{th:ordpa}.}
From our local approximation to the probability premium under DT,
we have
\begin{equation*}
\mu=-\frac{1}{2}\varepsilon_{1}^{2}\frac{h''(p_{0})}{h'(p_{0})}+o(\varepsilon_{1}^{2}),
\end{equation*}
which readily yields the first stated result.
Next, from our local approximation to the probability premium under RDU,
we have 
\begin{equation*}
\mu=-\frac{1}{2}\varepsilon_{1}^{2}\frac{h''(p_{0})}{h'(p_{0})}-\frac{1}{2}\varepsilon_{1}\varepsilon_{2}\frac{U''(w_{0})}{U'(w_{0})}
+o(\varepsilon_{1}^{2})+o(\varepsilon_{2}),
\end{equation*}
which yields the second stated result.
\qed

\noindent {\sc Proof of Theorem~\ref{th:ordpanondiff}.}
Recall the indifference relation that defines the DT probability premium:
\begin{equation*}
0=\left(h(p_{0}-\mu)-h(p_{0}-\varepsilon_{1})\right)\left(-1\right)+\left(h(p_{0}+\varepsilon_{1})-h(p_{0}-\mu)\right)\left(1\right),
\end{equation*}
hence
\begin{equation*}
0=h(p_{0}-\varepsilon_{1})-2h(p_{0}-\mu)+h(p_{0}+\varepsilon_{1}).
\end{equation*}
Then, upon differentiating both sides and taking the limit for $\varepsilon_{1}\downarrow 0$, we obtain
\begin{equation*}
0=-h'_{-}(p_{0})+2h'_{-}(p_{0})\frac{\partial \mu}{\partial \varepsilon_{1}}\Big|_{\varepsilon_{1}=0^{+}}+h'_{+}(p_{0}),
\end{equation*}
hence
\begin{align*}
\frac{\partial \mu}{\partial \varepsilon_{1}}\Big|_{\varepsilon_{1}=0^{+}}&=\frac{1}{2}\left(1-\frac{h'_{+}(p_{0})}{h'_{-}(p_{0})}\right)\\
&>(<)0,
\end{align*}
if $h'_{-}(p_{0})>(<) h'_{+}(p_{0})$.
\qed

\noindent {\sc Proof of Theorem~\ref{th:cra}.}
We first note that, due to Pratt \cite{P64}, Yaari \cite{Y86} and Eeckhoudt and Laeven \cite{EL20},
statement~(i) is equivalent to
\begin{itemize}
\item[(iii)] $\frac{U_{2}(y)-U_{2}(x)}{U_{2}(w)-U_{2}(v)}\leq \frac{U_{1}(y)-U_{1}(x)}{U_{1}(w)-U_{1}(v)}$
and
$\frac{h_{2}(s)-h_{2}(r)}{h_{2}(q)-h_{2}(p)}\leq \frac{h_{1}(s)-h_{1}(r)}{h_{1}(q)-h_{1}(p)}$
$\quad$ for all $v<w\leq x<y$ and all $0<p<q\leq r<s<1$.
\end{itemize}

Next, we will prove that the equivalent statements~(i) and (iii) imply statement~(ii).
Consider Eqn.~\eqref{eq:ppRDU} and fix an arbitrary $\varepsilon_{2}>0$.
Clearly, if $\varepsilon_{1}\rightarrow 0$ in \eqref{eq:ppRDU},
then $\mu_{i}\rightarrow 0$.
Define the function
\begin{align*}
W_{i}(\mu_{i},\varepsilon_{1})
:=&\left(h_{i}(p_{0}+\varepsilon_{1})-h_{i}(p_{0}-\varepsilon_{1})\right)U_{i}(w_{0})\nonumber\\
&-\left(\left(h_{i}(p_{0}-\mu_{i})-h_{i}(p_{0}-\varepsilon_{1})\right)U_{i}(w_{0}-\varepsilon_{2})\nonumber\right.\\
&\left.\quad+\left(h_{i}(p_{0}+\varepsilon_{1})-h_{i}(p_{0}-\mu_{i})\right)U_{i}(w_{0}+\varepsilon_{2})\right),\qquad i=1,2,
\end{align*}
and consider its total differential
\begin{equation*}
\mathrm{d}W_{i}=\frac{\partial W_{i}}{\partial \mu_{i}}\,\mathrm{d}\mu_{i} + \frac{\partial W_{i}}{\partial \varepsilon_{1}}\,\mathrm{d}\varepsilon_{1}.
\end{equation*}
It takes the form
\begin{align*}
\mathrm{d}W_{i}=-&h'_{i}(p_{0}-\mu_{i})\left(U_{i}(w_{0}+\varepsilon_{2})-U_{i}(w_{0}-\varepsilon_{2})\right)\,\mathrm{d}\mu_{i}\\
+&\left(h'_{i}(p_{0}-\varepsilon_{1})\left(U_{i}(w_{0})-U_{i}(w_{0}-\varepsilon_{2})\right)
-h'_{i}(p_{0}+\varepsilon_{1})\left(U_{i}(w_{0}+\varepsilon_{2})-U_{i}(w_{0})\right)\right)\,\mathrm{d}\varepsilon_{1}.
\end{align*}
We equate the total differential to zero.
We thus obtain
\begin{align}
\frac{\mathrm{d}\mu_{i}}{\mathrm{d}\varepsilon_{1}}=
\frac{U_{i}(w_{0})-U_{i}(w_{0}-\varepsilon_{2})}{U_{i}(w_{0}+\varepsilon_{2})-U_{i}(w_{0}-\varepsilon_{2})}\frac{h'_{i}(p_{0}-\varepsilon_{1})}{h'_{i}(p_{0}-\mu_{i})}
-\frac{U_{i}(w_{0}+\varepsilon_{2})-U_{i}(w_{0})}{U_{i}(w_{0}+\varepsilon_{2})-U_{i}(w_{0}-\varepsilon_{2})}\frac{h'_{i}(p_{0}+\varepsilon_{1})}{h'_{i}(p_{0}-\mu_{i})}.
\label{eq:tdpp}
\end{align}
Using statements~(i) and (iii) we will show that, for RDU DM $i=2$,
the first two fractions on the right-hand side of \eqref{eq:tdpp}
are larger than or equal to the corresponding fractions
for RDU DM $i=1$,
while the last two fractions are weakly smaller for DM $i=2$ than for DM $i=1$.
(All four fractions are positive for $i=1,2$.)

Indeed, first invoking statement~(i), we obtain, as in Pratt \cite{P64}, Eqn.~(20),
\begin{equation}
\frac{U_{2}'(x)}{U_{2}'(w)}\leq(\geq) \frac{U_{1}'(x)}{U_{1}'(w)},\qquad\mathrm{for}\ w<(>)x,
\label{eq:in(ii)}
\end{equation}
and we note that
the utility function $U_{i}$ can be replaced by the probability weighting function $h_{i}$
in \eqref{eq:in(ii)},
with the variables $w$ and $x$ 
restricted to $(0,1)$.
Next, statement~(iii) yields
\begin{equation*}
\frac{U_{2}(y)-U_{2}(x)}{U_{2}(w)-U_{2}(v)}+\frac{U_{2}(w)-U_{2}(v)}{U_{2}(w)-U_{2}(v)}
\leq \frac{U_{1}(y)-U_{1}(x)}{U_{1}(w)-U_{1}(v)}+\frac{U_{1}(w)-U_{1}(v)}{U_{1}(w)-U_{1}(v)},\qquad\mathrm{for}\ v<w\leq x<y.
\end{equation*}
In particular, when $w=x$ this reduces to
\begin{equation*}
\frac{U_{2}(y)-U_{2}(v)}{U_{2}(w)-U_{2}(v)}
\leq \frac{U_{1}(y)-U_{1}(v)}{U_{1}(w)-U_{1}(v)},\qquad\mathrm{for}\ v<w<y,
\end{equation*}
whence
\begin{equation}
\frac{U_{2}(w)-U_{2}(v)}{U_{2}(y)-U_{2}(v)}
\geq \frac{U_{1}(w)-U_{1}(v)}{U_{1}(y)-U_{1}(v)}, 
\label{eq:in(iii)-i}
\end{equation}
and
\begin{equation}
\frac{U_{2}(y)-U_{2}(w)}{U_{2}(y)-U_{2}(v)}
\leq \frac{U_{1}(y)-U_{1}(w)}{U_{1}(y)-U_{1}(v)}.
\label{eq:in(iii)-ii}
\end{equation}

Invoking the inequalities 
in Eqn.~\eqref{eq:in(ii)}
(with $U_{i}$ replaced by $h_{i}$ upon restricting the corresponding domains to $(0,1)$)
and Eqns.~\eqref{eq:in(iii)-i} and \eqref{eq:in(iii)-ii},
we find from \eqref{eq:tdpp} that
\begin{equation}
\frac{\mathrm{d}\mu_{2}}{\mathrm{d}\varepsilon_{1}}\geq \frac{\mathrm{d}\mu_{1}}{\mathrm{d}\varepsilon_{1}},
\label{eq:tdpp2}
\end{equation}
hence (ii).

We have thus shown that statement~(i) implies statement~(ii).
It remains to show that (ii) also implies (i),
or, equivalently, that not (i) implies not (ii).
To this end, it suffices to note that
if statement~(i) fails to hold on some interval (of $x$ or $p$),
then,
due to the arbitrariness of $w_{0}$, $p_{0}$, $\varepsilon_{2}>0$ and $\varepsilon_{1}$ with
$0<\varepsilon_{1}\leq\{p_{0},1-p_{0}\}<1$,
one can always find feasible $w_{0}$, $p_{0}$, $\varepsilon_{2}$ and $\varepsilon_{1}$ such that
\eqref{eq:tdpp2} 
holds on some interval with the inequality sign strict and reversed,
and as a consequence statement~(ii) fails to hold.
This completes the proof.
\qed

\setcounter{equation}{0}

\section{Generalization to $n$-State Risks
}\label{sec:nonbinary}

Let $n\geq 2$ be an integer.
Consider an $n$-state risk with total probability mass equal to $2\varepsilon_{1}$
assigning a probability of $\tfrac{2\varepsilon_{1}}{n}$ to the payoffs $x_{i}$, $i=1,\ldots,n$,
with $x_{1}\leq\cdots\leq x_{n}$,
where $0<\varepsilon_{1}\leq\tfrac{1}{2}$.
The increments between the $x_{i}$'s may equal zero;
the 
risk would then feature unequal state probabilities.
We denote by $n_{1}$ the number of states with negative payoff and let $n_{2}=n-n_{1}$.
Note the generality of this setting.

We suppose that $\sum_{i=1}^{n}x_{i}=0$, so the risk has mean equal to zero.
We also suppose that $|x_{i}|<\varepsilon_{2}$ for some $\varepsilon_{2}>0$, so $\sum_{i=1}^{n} x_{i}^{2}=O(\varepsilon_{2}^{2})$.

One easily verifies that the first and second primal and dual moments
are given by
\begin{align}
\texttt{m}_{1}&=\frac{2\varepsilon_{1}}{n}\sum_{i=1}^{n}x_{i}=0,\label{eq:firstprimaldual}\\
\texttt{m}_{2}&=\frac{2\varepsilon_{1}}{n}\sum_{i=1}^{n}x_{i}^{2},\label{eq:secondprimal}\\
\bar{\texttt{m}}_{2}&=\frac{4\varepsilon_{1}^{2}}{n^{2}}\sum_{i=1}^{n}(2i-1)x_{i}.\label{eq:seconddual}
\end{align}
Note that $\texttt{m}_{2}=O(\varepsilon_{2}^{2})$
and that $\bar{\texttt{m}}_{2}=O(\varepsilon_{1}^{2})$.

In this general setting, we define the probability premium of the zero-mean $n$-state risk
as the (uniform) reduction in the probabilities of the unfavorable (i.e., negative payoff) states,
inducing a (uniform) increase in the probabilities of the favorable states accordingly,
such that the DM is indifferent between bearing and not bearing the risk.
Note that this definition of the probability premium applies
irrespective of whether $\varepsilon_{1}$ and $\varepsilon_{2}$ are ``small'' or ``large''.

Formally, the RDU probability premium, $\mu$, thus solves
\begin{align*}
&
\left(h(p_{0}+\varepsilon_{1})-h(p_{0}-\varepsilon_{1})\right)U\left(w_{0}\right)
\\
&=
\sum_{i=1}^{n_{1}}\left(h\left(p_{0}-\varepsilon_{1}+\frac{2\varepsilon_{1}}{n}i-\mu i\right)
-h\left(p_{0}-\varepsilon_{1}+\frac{2\varepsilon_{1}}{n}(i-1)-\mu(i-1)\right)\right)U\left(w_{0}+x_{i}\right)\\
&\quad
+\sum_{i=n_{1}+1}^{n}\left(h\left(p_{0}-\varepsilon_{1}-\mu n_{1}+\frac{2\varepsilon_{1}}{n}i+\mu\frac{n_{1}}{n_{2}}(i-n_{1})\right)\right.\\
&\qquad\qquad\qquad
\left.-h\left(p_{0}-\varepsilon_{1}-\mu n_{1}+\frac{2\varepsilon_{1}}{n}(i-1)+\mu\frac{n_{1}}{n_{2}}(i-1-n_{1})\right)\right)U\left(w_{0}+x_{i}\right).
\end{align*}

Invoking second-order Taylor series expansions of 
$U$ around $w_{0}$
and of 
$h$ around $p_{0}$
yields, upon rearranging terms,
\begin{align*}
0&\simeq
\sum_{i=1}^{n_{1}}\left(h'\left(p_{0}\right)\left(\frac{2\varepsilon_{1}}{n}-\mu\right)\right.\\
&\left.\qquad\qquad+\frac{1}{2}h''\left(p_{0}\right)
\left(\left(-\varepsilon_{1}+\frac{2\varepsilon_{1}}{n}i-\mu i\right)^{2}
-\left(-\varepsilon_{1}+\frac{2\varepsilon_{1}}{n}(i-1)-\mu(i-1)\right)^{2}\right)\right)\\
&\quad\qquad\times\left(U'\left(w_{0}\right)x_{i}+\frac{1}{2}U''\left(w_{0}\right)x_{i}^{2}\right)\\
&+\sum_{i=n_{1}+1}^{n}\left(h'\left(p_{0}\right)\left(\frac{2\varepsilon_{1}}{n}+\mu\frac{n_{1}}{n_{2}}\right)\right.\\
&\left.\qquad\qquad+\frac{1}{2}h''\left(p_{0}\right)
\left(\left(-\varepsilon_{1}-\mu n_{1}+\frac{2\varepsilon_{1}}{n}i+\mu\frac{n_{1}}{n_{2}}(i-n_{1})\right)^{2}\right.\right.\\
&\qquad\qquad\qquad\qquad\qquad\left.\left.-\left(-\varepsilon_{1}-\mu n_{1}+\frac{2\varepsilon_{1}}{n}(i-1)+\mu\frac{n_{1}}{n_{2}}(i-1-n_{1})\right)^{2}\right)\right)\\
&\quad\qquad\times\left(U'\left(w_{0}\right)x_{i}+\frac{1}{2}U''\left(w_{0}\right)x_{i}^{2}\right).
\end{align*}

Hence, upon further rearranging terms and using $\sum_{i=1}^{n}x_{i}=0$,
we obtain at the leading orders in $\varepsilon_{1}$ and $\varepsilon_{2}$,
the local approximation
\begin{align*}
\mu&\simeq -\frac{1}{2}\frac{\frac{2\varepsilon_{1}}{n}\sum_{i=1}^{n}x_{i}^{2}}{\sum_{i=1}^{n_{1}}|x_{i}|+\sum_{i=n_{1}+1}^{n}\tfrac{n_{1}}{n_{2}}|x_{i}|}\frac{U''(w_{0})}{U'(w_{0})}
-\frac{1}{2}\frac{\frac{4\varepsilon_{1}^{2}}{n^{2}}\sum_{i=1}^{n}(2i-1)x_{i}}{\sum_{i=1}^{n_{1}}|x_{i}|+\sum_{i=n_{1}+1}^{n}\tfrac{n_{1}}{n_{2}}|x_{i}|}\frac{h''(w_{0})}{h'(w_{0})}\\
&=\frac{\texttt{m}_{2}}{2\texttt{Py}^{\ast}}\left(-\frac{U''(w_{0})}{U'(w_{0})}\right)
+\frac{\bar{\texttt{m}}_{2}}{2\texttt{Py}^{\ast}}\left(-\frac{h''(p_{0})}{h'(p_{0})}\right),
\end{align*}
using \eqref{eq:secondprimal}--\eqref{eq:seconddual} in the last equality
and where $\texttt{Py}^{\ast}=\sum_{i=1}^{n_{1}}|x_{i}|+\sum_{i=n_{1}+1}^{n}\tfrac{n_{1}}{n_{2}}|x_{i}|$,
the weighted $1$-norm of the payoffs of the $n$-state risk.
Of course, when $n_{1}\equiv n_{2}$, $\texttt{Py}^{\ast}=\sum_{i=1}^{n}|x_{i}|=\texttt{Py}$,
the standard $1$-norm.

\end{appendices}

\newpage

\vskip 1 cm

\textbf{Acknowledgements.}\
We are very grateful
to
Lo\"ic Berger 
and 
Christian Gollier for 
comments
and suggestions and to Harris Schlesinger ($\dag$) for discussions.
This research was funded in part by
the French Agence Nationale de la Recherche under grant ANR-17-CE03 (Eeckhoudt, project Induced)
and the Netherlands Organization for Scientific Research under grants NWO VIDI 2009 and NWO VICI 2019/2020 (Laeven).
Research assistance of Andrei Lalu is gratefully acknowledged.
Part of this paper was circulated earlier under the title
``Risk Aversion in the Small and in the Large under Rank-Dependent Utility''.

\begin{spacing}{0.0}

\end{spacing}

\end{document}